\documentclass[aip,pop,reprint,amsmath,amssymb,onecolumn]{revtex4-1}

\usepackage[dvipdfmx]{graphicx}% Include figure files
\usepackage{color}
\renewcommand{\vec}[1]{\boldsymbol{#1}}

\def\apj{{\itshape Astrophys. J.} }
\def\jgr{{\itshape J. Geophys. Res.} }
\def\pop{{\itshape Phys. Plasmas} }

\begin{document}

\title{An approximate Kappa generator for particle simulations}

\author{Seiji Zenitani}
\affiliation{Space Research Institute, Austrian Academy of Sciences, 8042 Graz, Austria}
\email{seiji.zenitani@oeaw.ac.at}
\author{Takayuki Umeda}
\affiliation{Information Initiative Center, Hokkaido University, Sapporo 060-0811, Japan}

\begin{abstract}
A random number generator for the Kappa velocity distribution in particle simulations is proposed.
Approximating the cumulative distribution function with the $q$-exponential function, 
an inverse transform procedure is constructed.
The proposed method provides practically accurate results,
in particular for $\kappa < 4$.
It runs fast on graphics processing units (GPUs).
The derivation, numerical validation, and relevance to GPU execution models are discussed. 
\end{abstract}


\maketitle

\section{Introduction}

In space physics,
the Kappa distribution is arguably the most important velocity distribution
after the Maxwell distribution.
Since it was empirically introduced during the 1960s \citep{olbert68,vas68}, the Kappa distribution has been extensively observed throughout the heliosphere \citep{pierrard10,kappa} and it has become increasingly popular.
Theoretically, it has also been established that the Kappa distribution is related to Tsallis statistics \citep{livadiotis09}.
Since the Kappa distribution maximizes non-extensive entropy,
it may represent a terminal state after plasma long-range interactions \citep{tsallis23}.
Driven by spacecraft observations and various theoretical motivations,
the kinetic modeling of Kappa-distributed plasmas is growing in its importance \citep{park13,gedalin22,fitzmaurice24,nechaev25}.

In kinetic plasma simulations,
it is often necessary to generate particle velocity distributions
by using random numbers (random variates) at the initialization stage.
It is straightforward to produce the Maxwell distribution,
because one can employ a random number generator (RNG)
for the normal (Gaussian) distribution,
which is built into numerous programming languages and libraries.
One can also implement the Box--Muller method
for generating two independent normal variates from two uniform variates \citep{bm58}.
Typically, three independent normal variates per particle are necessary to initialize a Maxwellian plasma.
Related to this, one of the authors has developed
an inverse sampling method for the 3D Maxwell distribution \citep{umeda24}.
The author approximated
the cumulative distribution function of the Maxwellian
by using an invertible function, and then
constructed an inverse transform procedure.

The standard Kappa distribution can be obtained
by dividing three normal variates by a single gamma variate \citep{abdul15,zeni22}.
This is similar to an RNG for Student's $t$-distribution,
because the Kappa distribution is a multivariate $t$-distribution.
It requires a gamma generator. In case it is unavailable,
one can implement one of many gamma generators \citep{devroye86,luengo22}.
Alternatively, one can generate
the standard Kappa distribution by using an acceptance--rejection method
from the Type II Pareto distribution \citep{zeni25}.
Importantly, most gamma generators internally employ
the rejection method.

Due to their excellent performance per watt (computational power per electric power consumption), high-performance computing on Graphics Processing Units (GPUs)
has become increasingly important. 
Typical GPUs employ the Single-Instruction, Multiple-Threads (SIMT) execution model, in which a single instruction controls a group of threads simultaneously.
This model is efficient at executing the same calculation in parallel;
however, it is not good at handling loops
whose iteration counts differ from thread to thread. 
This is the case for the rejection method.
This may be one reason
why numerical libraries on GPUs such as {\tt cuRAND} and {\tt hipRAND}
do not provide gamma RNGs,
even though they offer uniform and normal RNGs.

Because of a growing demand for kinetic modeling and ongoing transition to GPU-based computers,
we desire a GPU-friendly RNG for the Kappa distribution. 
In this study, we propose a novel Kappa generator,
inspired by our previous work on the Maxwell distribution \citep{umeda24}.
Approximating the cumulative distribution function with an invertible function,
we propose an inverse transform method. 
The algorithm is GPU-friendly, because it has no control flow divergence.
It generates an approximate distribution
that is indistinguishable from the Kappa distribution unless a very large number of particles are used.

The remainder of this paper is organized as follows.
Section 2 presents the standard Kappa distribution.
Section 3 derives our approximate Kappa distribution,
starting from the CDF of the Kappa distribution.
Section 4 constructs the inverse transform procedure.
Section 5 evaluates the accuracy and performance of the new method by benchmarks.
Section 6 discusses the results with respect to the SIMT execution model, and then concludes this paper.
%Section \ref{sec:kappa} presents the standard Kappa distribution.
%Section \ref{sec:approx} derives our approximate Kappa distribution,
%starting from the CDF of the Kappa distribution.
%Section \ref{sec:generator} constructs the inverse transform procedure.
%Section \ref{sec:test} evaluates the accuracy and performance of the new method by benchmarks.
%Section \ref{sec:discussion} discusses the results with respect to the SIMT execution model, and then concludes this paper.

\section{Kappa distribution}
\label{sec:kappa}

The phase space density of the (nonrelativistic) Kappa distribution is
\begin{align}
\label{eq:kappa}
f(\vec{v})  d^3v
= \frac{N_{\kappa}}{(\pi\kappa\theta^2)^{3/2}} \frac{\Gamma(\kappa+1)}{\Gamma(\kappa-1/2)}  \Big( 1 + \frac{ \vec{v}^2 }{\kappa \theta^2} \Big)^{-(\kappa+1)} d^3v
\end{align}
where $N_{\kappa}$ is the number density, $\theta$ is the most probable speed, $\Gamma(x)$ is the gamma function, and $\kappa > 3/2$ is the kappa index. 
In this study, we discuss the Kappa distribution as a probability distribution function; thus, we hereafter set $N_{\kappa}=1$. 
Introducing a new variable $x \equiv v^2/\theta^2$
in the polar coordinates ($d^3v \rightarrow 4\pi v^2 dv$),
we translate the Kappa distribution into
a beta-prime distribution, also known as the beta distribution of the second kind, with shape $(3/2,\kappa-1/2)$:
\begin{align}
f(x) dx
&=
\frac{1}{\kappa\,B\left( \frac{3}{2},\kappa - \frac{1}{2} \right)} 
\left(\frac{x}{\kappa}\right)^{1/2}
\Big( 1 + \frac{x}{\kappa} \Big)^{-(\kappa+1)} dx
\label{eq:kappax}
\end{align}
Here, $B(x,y)$ is the beta function. 
Note that Eq.~\eqref{eq:kappax} contains a scaling factor $\kappa$.
The cumulative distribution function (CDF) of this distribution yields
\begin{align}
F(x)
&= 
I_{\frac{x}{x+\kappa}}\left( \frac{3}{2}, \kappa-\frac{1}{2} \right)
\label{eq:CDF}
\end{align}
where $I_x$ is the regularized incomplete beta function. 
Similarly, the distribution of the energy density yields
\begin{align}
f_E(x) dx
&=
f(x) \cdot \frac{1}{2}mx\theta^2 \,dx
\nonumber\\
&=
\frac{3\kappa}{2(2\kappa-3)} m\theta^2
%\nonumber\\
%&~~~\times
\left\{
\frac{1}{\kappa B\left( \frac{5}{2},\kappa - \frac{3}{2} \right)} 
\left(\frac{x}{\kappa}\right)^{3/2}
\Big( 1 + \frac{x}{\kappa} \Big)^{-(\kappa+1)}
\right\}\,dx
\end{align}
where the term in curly braces indicates the beta-prime distribution with shape $(5/2,\kappa-3/2)$.
Then we obtain the total energy density
\begin{align}
\label{eq:kappaE}
\mathcal{E}_{\kappa}
= \int_0^{\infty}f_E(x)dx
= \frac{3\kappa}{2(2\kappa-3)} m\theta^2
\end{align}
as well as the CDF of the energy density distribution
\begin{align}
F_E(x)
&= 
I_{\frac{x}{x+\kappa}}\left( \frac{5}{2}, \kappa-\frac{3}{2} \right)
\label{eq:CDF_E}
\end{align}

\section{Approximate Kappa distribution}
\label{sec:approx}

In line with \citet{umeda24},
we introduce the following approximation of the CDF:
\begin{align}
F(x) \approx G(x) \equiv
\left\{ 1 - \exp_{q*}\left( -\frac{ax+bx^2}{1+cx} \right) \right\}^{3/2}
.\label{eq:G}
\end{align}
Here,
\begin{align}
\exp_q(x) \equiv \bigg( 1 + (1-q) x \bigg)^{\frac{1}{1-q}}
\label{eq:qexp}
\end{align}
is the $q$-exponential function with the entropy index $q$.
We use fitting parameters $q^* \equiv 1+1/\kappa^*$ and $\kappa^*$,
because the entropy index is expected to be $q=1+1/\kappa$ in Kappa distribution theory. 
Eq.~\eqref{eq:G} has four parameters: $a, b, c,$ and $\kappa^* (q^*)$.

By differentiating $G(x)$,
we obtain approximate distribution functions in the $x$ space and in the velocity space.
\begin{align}
g(x) dx
&=
\dfrac{3 (a + 2bx + bcx^{2})}{2\left(1 + cx \right)^{2}}
%\nonumber\\ &~~\times
\sqrt{1 - \left( 1 + \frac{ax + bx^{2}}{\kappa^*(1 + cx)} \right)^{-\kappa^*} }
%\nonumber\\ &~~\times
\left( 1 + \frac{ax + bx^{2}}{\kappa^*(1 + cx)} \right) ^{-(\kappa^*+1)}
dx
\label{eq:approx_x}
\\
g(\vec{v}) d^3v
&=
\dfrac{3 (a\theta^4 + 2bv^2\theta^2 + bcv^4)}{4\pi v \theta^2 \left(\theta^2 + cv^2 \right)^{2}}
%\nonumber\\ &~~\times
\sqrt{1 - \left( 1 + \frac{(a\theta^2 + bv^2) v^2}{\kappa^*(\theta^2 + cv^2) \theta^2} \right)^{-\kappa^*} }
%\nonumber\\ &~~\times
\left( 1 + \frac{(a\theta^2 + bv^2)v^2}{\kappa^*(\theta^2 + cv^2)\theta^2} \right) ^{-(\kappa^*+1)}
d^3v
\label{eq:approx}
\end{align}

\begin{figure}[htbp]
\centering
\includegraphics[width={0.6\columnwidth}]{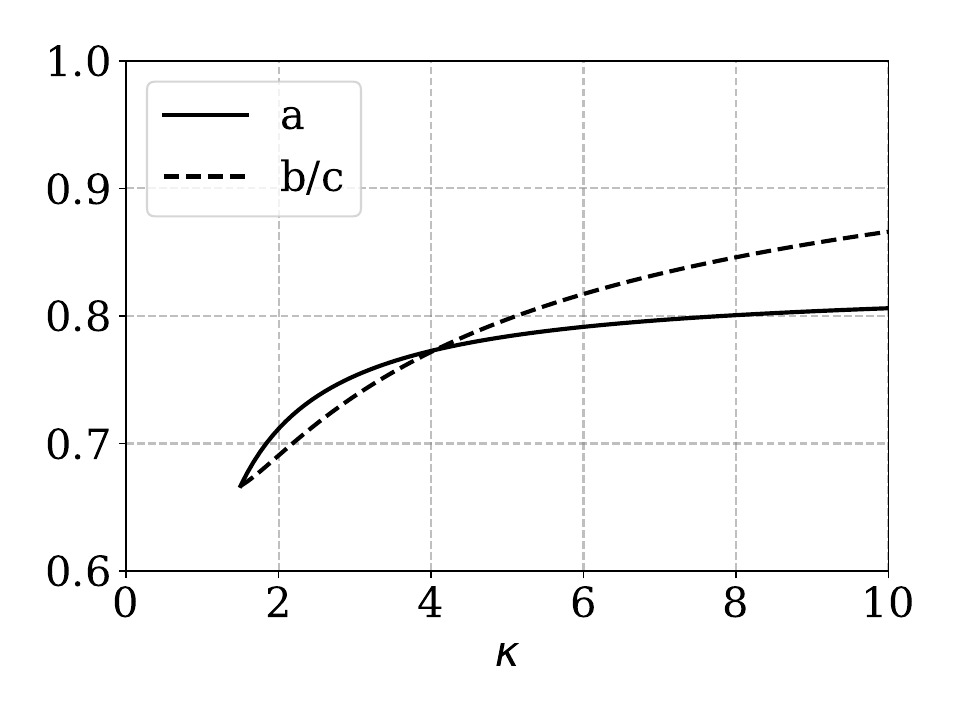}
\caption{
The two parameters, $a$ (Eq.~\eqref{eq:a}) and $b/c$ (Eq.~\eqref{eq:b}), are displayed
as functions of $\kappa$.
\label{fig:curve}}
\end{figure}

Next, we seek the four parameters.
We consider Taylor expansions of Eqs.~\eqref{eq:CDF} and \eqref{eq:G} near $x=0$.
\begin{align}
F(x) &=
\frac{2}{3 B\left( \frac{3}{2},\kappa - \frac{1}{2} \right)}
\left(\frac{x}{\kappa}\right)^{3/2}
\left(
1
-
\frac{3(\kappa+1)}{5\kappa}x
+
\mathcal{O}(x^2)
\right)
\label{eq:F_taylor_0}
\\
G(x) &=
a^{3/2}
x^{3/2}
\left\{
1
+
\frac{3}{2}
\left(
\frac{b-ac}{a}
-
\frac{(\kappa^*+1)a}{2\kappa^*}
\right)
x
+
\mathcal{O}(x^2)
\right\}
\label{eq:G_taylor_0}
\end{align}
Their derivations are given in the Appendix.
Comparing the coefficients of the $x^{3/2}$ terms,
we find
\begin{align}
a
=
\frac{1}{\kappa}
\left(\frac{2}{3 B\left( \frac{3}{2},\kappa - \frac{1}{2} \right)}\right)^{2/3}
\label{eq:a}
\end{align}
Similarly comparing Taylor expansions for $x\rightarrow\infty$
(also derived in the Appendix),
\begin{align}
F(x) &=
1 - 
\frac{\kappa^{\kappa-1/2}}{(\kappa-1/2)B\left( \frac{3}{2},\kappa - \frac{1}{2} \right)}
x^{-(\kappa-1/2)}
+
\mathcal{O}(x^{-(\kappa+1/2)})
\label{eq:F_taylor_8}
\\
G(x) &=
1
-
\frac{3}{2}
\left(\frac{c \kappa^*}{b}\right)^{\kappa^*}
x^{-\kappa^*}
+
\mathcal{O}(x^{-(\kappa^*+1)})
\label{eq:G_taylor_8}
\end{align}
we find
\begin{align}
\kappa^* &= \kappa -\frac{1}{2},
\label{eq:k2}
\\
\frac{b}{c}
&=
\left\{ \kappa^* \frac{3}{2} B\left( \frac{3}{2},\kappa - \frac{1}{2} \right)\right\}^{1/\kappa^*} \frac{\kappa^*}{\kappa}
\label{eq:b}
\end{align}
The two parameters $a$ and $b/c$ are shown in Figure \ref{fig:curve}
as functions of $\kappa$. 
Starting from $2/3$ at $\kappa=3/2$, they increase monotonically,
intersect each other at $\kappa \approx 4.1$, and then approach to their asymptotic limits.
In the $\kappa \rightarrow \infty$ limit, considering $B(\alpha,\beta) \approx \Gamma(\alpha)\beta^{-\alpha}$ for a large $\beta$, we find
\begin{align}
a \rightarrow  \left( \frac{4}{3\sqrt{\pi}} \right)^{2/3} \approx 0.827
.
\end{align}
We also find $b/c \rightarrow 1 $ for $\kappa \rightarrow \infty$.

\begin{figure}[htbp]
\centering
\includegraphics[width={0.7\columnwidth}]{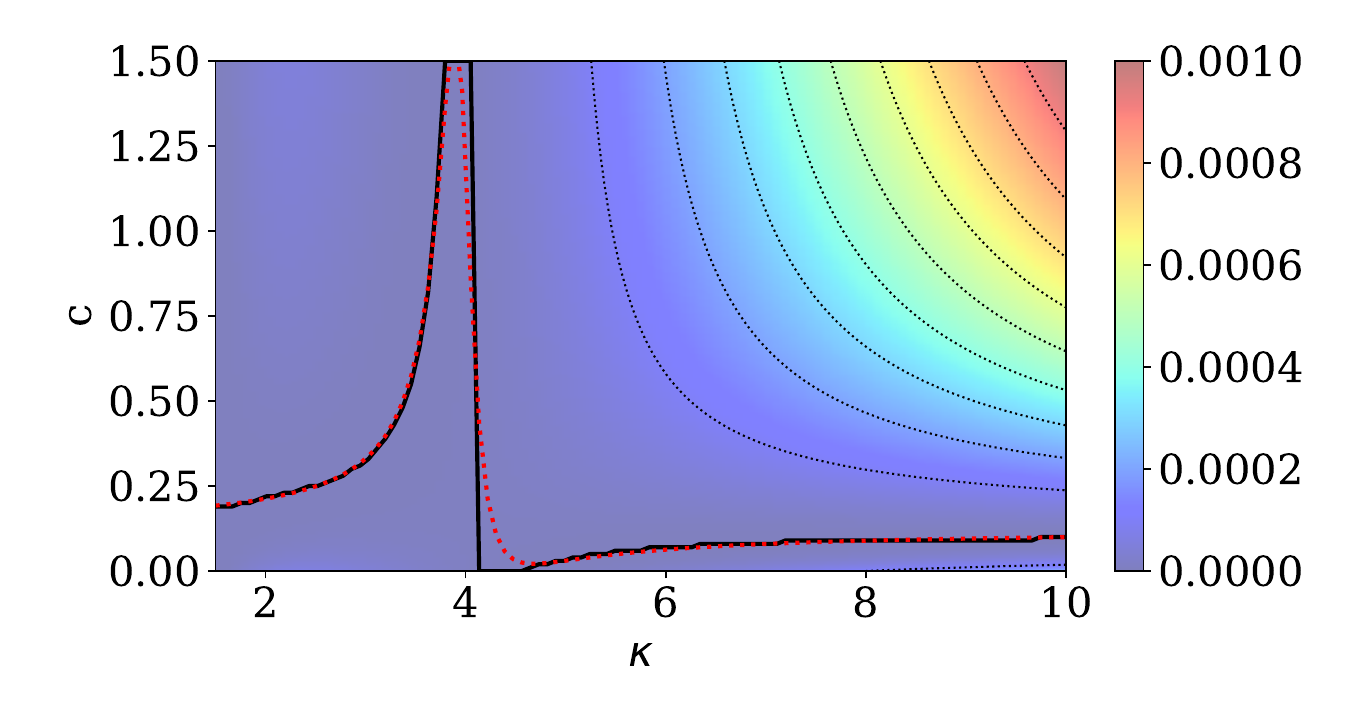}
\caption{
The relative entropy $\mathcal{D}$ (colormap) is shown as a function of $(\kappa,c)$.
The optimum line (black) and the approximate line (Eq.~\eqref{eq:c}; red) are overplotted.
\label{fig:map}}
\end{figure}

To determine the last parameter $c$, we use the grid search method.
We evaluate the relative entropy $\mathcal{D}$,
also known as the Kullback--Leibler (KL) divergence, between
the Kappa distribution (Eq.~\eqref{eq:kappax}) and our approximation (Eq.~\eqref{eq:approx_x}):
\begin{align}
\mathcal{D}(f\|g; \kappa, c)
&= \int_0^{\infty} f(x;\kappa) \ln \frac{f(x;\kappa)}{g(x;\kappa, c)} dx
%\\
%&= \int_0^{\infty} f(x;\kappa) \ln f(x;\kappa) dx - H(f,g ; \kappa)
\label{eq:relative_entropy}
\end{align}
It approaches zero when the two distributions are similar,
while it increases when the target probability distribution $g(x)$ deviates
from the baseline probability distribution $f(x)$.
We have carried out a survey of $\mathcal{D}$
in the 2D parameter space of $c\in [0,1.5]$ and $\kappa \in [1.5,10]$.
Then, for a given $\kappa$, we look for $c$ that minimizes $\mathcal{D}$.
Figure \ref{fig:map} shows $\mathcal{D}$ as a function of $(\kappa,c)$.
The black curve shows an optimum $c$ that minimizes $\mathcal{D}$.
The optimum $c$ diverges to infinity,
before it returns to zero at $\kappa\approx 4.1$.
This suggests that
an optimum value of $(ax+bx^2)/(1+cx)$ is $\approx bx/c$ in the left vicinity of $\kappa\approx 4.1$ and $\approx ax$ in the right vicinity.
This fractional term is continuous,
because the two factors $a$ meets $b/c$ at $\kappa\approx 4.1$
(Fig.~\ref{fig:curve}). 
The parameter $c$ essentially controls
the weights of $bx/c$ and $ax$ inside the fraction equation. 
When the two factors $a$ and $b/c$ are similar,
it is less meaningful to choose $c$ near $\kappa\approx 4.1$. 
%In other words, since $(ax+bx^2)/(1+cx) = (ax+c[b/c]x^2)/(1+cx)$,
%$c$ is 
%when $a = b/c$,
%becomes similar regardless of $c$.
Considering this, we look for a mathematical function
that seamlessly connects the optimum $c$ for $\kappa \ll 4$ and for $\kappa \gg 4$.
Using a symbolic regression library {\tt PySR} \citep{pysr},
we find the following approximate formula from
the optimum $c$ dataset in $\kappa \in [1.5, 3.6]$ and $[4.6, 10.0]$, and
an artificial anchor point of $(\kappa,c) = (4.1, 0.4)$.
\begin{align}
c_{\rm approx} &= \frac{0.123 \kappa^{2} - 1.12\, \kappa + 2.56}{\kappa^{2} - 7.89\, \kappa + 15.6}
\label{eq:c}
\end{align}
This solution is shown in the red dotted line in Figure \ref{fig:map}.
It hits its maximum $\approx 1.6$ at $\kappa \approx 3.9$ and
its minimum $\approx 0.02$ at $\kappa \approx 4.7$.

%\begin{align}
%x = (v/\theta)^2 = \frac{-(a+cL) + \sqrt{(a+cL)^2 - 4bL} }{2b}
%\end{align}

\begin{figure*}[htbp]
\centering
\includegraphics[width={0.9\textwidth}]{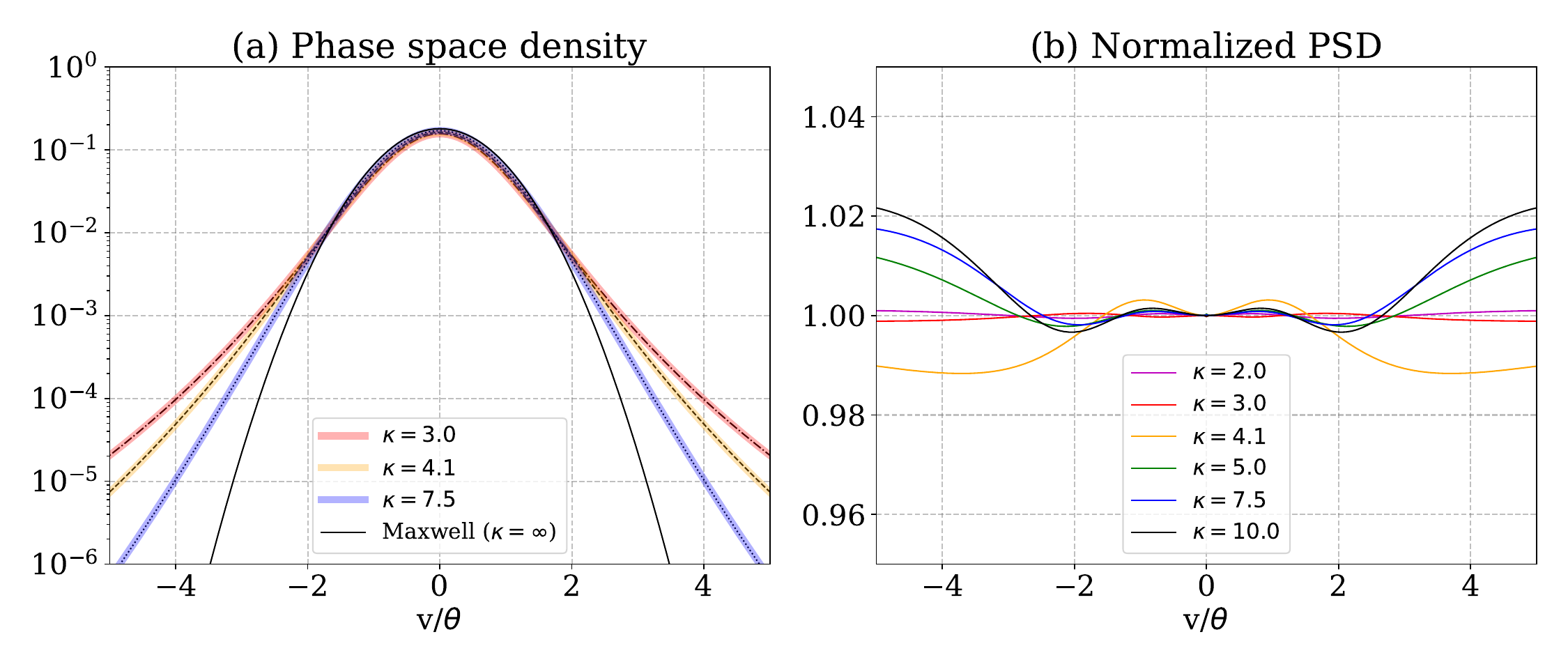}
\caption{
Comparison of the Kappa distribution (Eq.~\eqref{eq:kappa}) and the proposed approximation (Eq.~\eqref{eq:approx}).
(a) Phase space density as a function of $v$.
The thin black lines indicate the Kappa distribution, and
the color thick lines indicate the proposed approximation.
(a) Phase space density of the proposed approximation, normalized by the exact Kappa distribution.
\label{fig:PSD}}
\end{figure*}

Now we have all the four parameters (Eqs.~\eqref{eq:a}, \eqref{eq:b}, \eqref{eq:c}, and \eqref{eq:k2}).
Let us see the properties of the approximate distribution.
Figure \ref{fig:PSD} compares the original Kappa distribution (Eq.~\eqref{eq:kappa}) and
the proposed approximation (Eq.~\eqref{eq:approx}).
Hereafter, the number density is set to $N_{\kappa}=1$.
Figure \ref{fig:PSD}(a) displays
phase space density (PSD) of the Kappa distribution (the thin black lines)
and the proposed approximation (the color thick lines).
They are in excellent agreement, at least on the logarithmic scale.
To see the difference,
we compare the ratio of PSDs in Figure \ref{fig:PSD}(b).
We find that the approximated distribution is remarkably accurate
for $\kappa \lesssim 3$.
As $\kappa$ increases, there are deviations from the original distribution
in the $v \gg \theta$ region in Figure \ref{fig:PSD}(b).
Fortunately, the absolute error is small,
because the PSD is already very small,
in particular when $\kappa$ is large.
Considering that
the Kappa distribution ($4\pi v^2 f(\vec{v})$) has
maximum population at $v=\theta$,
it is reasonable that 
Eq.~\eqref{eq:approx} provides a good approximation
near $|v|=\theta$
for all values of $\kappa$ in Figure \ref{fig:PSD}(b). 

\begin{figure*}[htbp]
\centering
\includegraphics[width={0.9\textwidth}]{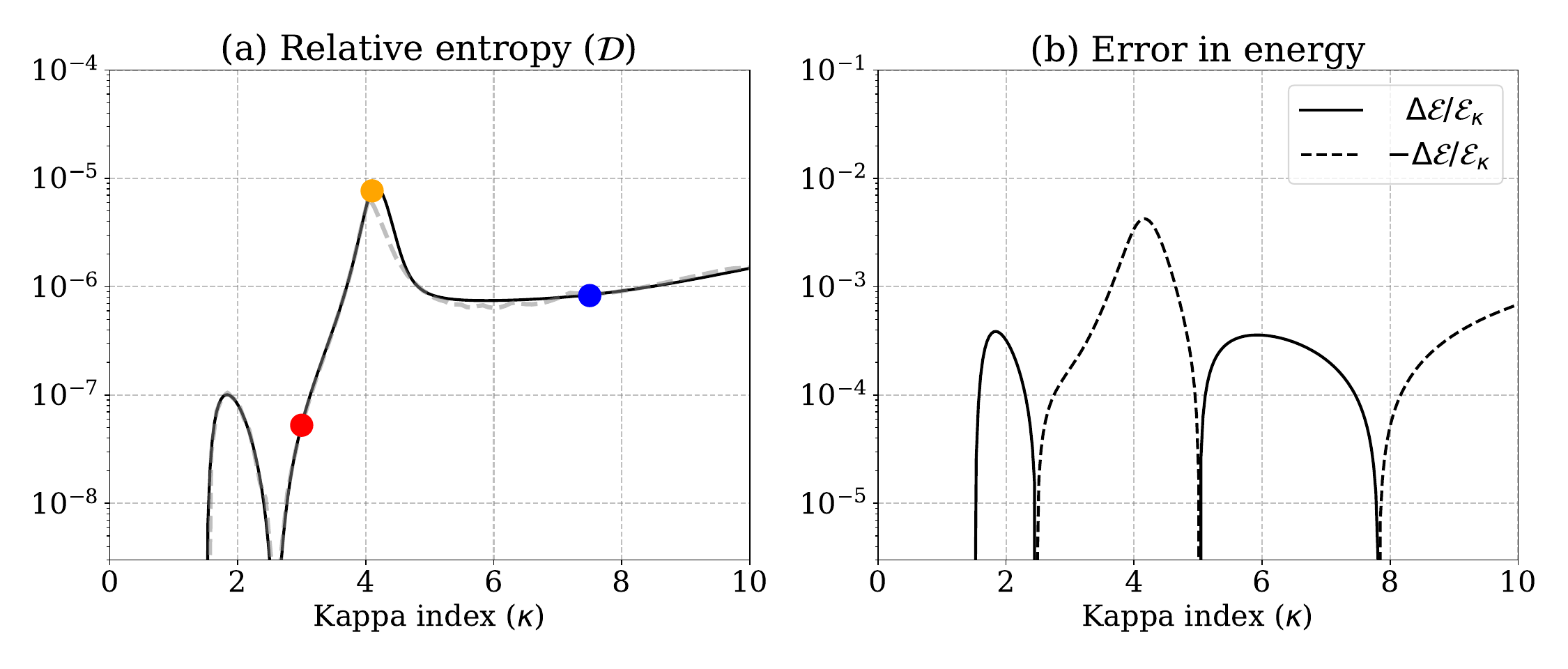}
\caption{
(a) The relative entropy between the approximate distribution (Eq.~\eqref{eq:approx_x}) and the exact distribution (Eq.~\eqref{eq:kappax}) as a function of $\kappa$.
(b) The normalized difference in energy density, $\pm \Delta \mathcal{E} / \mathcal{E}_{\kappa} = \pm (\mathcal{E}_{\rm approx}-\mathcal{E}_{\kappa}) / \mathcal{E}_{\kappa}$ as a function of $\kappa$.
\label{fig:DE}}
\end{figure*}

Figure \ref{fig:DE}(a) shows
the relative entropy $\mathcal{D}$ of our approximation.
This corresponds to the relative entropy along the approximate line (the red dashed line) in Figure \ref{fig:map}, $\mathcal{D}(f\|g; \kappa, c_{\rm approx})$. 
For reference, the relative entropy along the optimum line (the black line in Fig.~\ref{fig:map}) is also presented by the gray dashed line.
If we take a closer look, the optimum case looks somewhat jagged and
it even surpasses the approximate case.
These are due to the limited resolution of our grid search. 
Aside from these, the two entropies look similar,
suggesting that the handy function (Eq.~\eqref{eq:c}) works excellently. 
The relative entropy is remarkably low for $\kappa < 4$.
This suggests that the two distributions are similar.
It drops near $\kappa \approx 2.6$, but remains non-zero, 
because there is a subtle difference between the two distributions. 
The relative entropy has a peak at $\kappa \approx 4.2$.
For $\kappa \gtrsim 5$, $\mathcal{D}$ drops to $\approx 10^{-6}$, and then
it starts to increase very slowly. 
The three circles indicate values for $\kappa=3.0, 4.1,$ and $7.5$,
corresponding to the three representative cases in Figure \ref{fig:PSD}(a).
They effectively represent the three regimes of
low-$\kappa$, high-$\kappa$, and the transition between them.

The total density of the approximate distribution is,
by definition, $N_{\kappa}$.
Let us evaluate the total energy density of the distribution,
$\mathcal{E}_{\rm approx}$.
\begin{align}
\mathcal{E}_{\rm approx} = \int_0^{\infty} g(x;\kappa) \left( \frac{1}{2}mv^2 \right) dx
\end{align}
We numerically calculate $\mathcal{E}_{\rm approx}$, and then
we evaluate the difference to
that of the Kappa distribution (Eq.~\eqref{eq:kappaE}).
Figure~\ref{fig:DE}(b) shows the relative error in energy,
$\pm {\Delta \mathcal{E}}/{\mathcal{E}_{\kappa}} = \pm ({\mathcal{E}_{\rm approx} - \mathcal{E}_{\kappa}})/\mathcal{E}_{\kappa}$, and
the positive and negative errors are shown by the solid and dashed lines.
The error sometimes becomes positive, and sometimes negative.
It becomes zero at $\kappa \approx 2.5$,
which is slightly different from the minimum entropy point.
The energy density matches that of the Kappa distribution
within an error of $10^{-3}$,
except for the largest error at $\kappa \approx 4.1$.
Let us evaluate the anomaly at $\kappa \approx 4.1$ ---
from Figure \ref{fig:PSD}(b), we see that
plasma density is underestimated by 1\%
in the high-energy region of $v > 3\theta$.
This suggests that 1\% of the tail population in $v > 3\theta$
is erroneously lost and redistributed in the low-energy region.
From Eq.~\eqref{eq:CDF_E} with $x=3^2$ and $\kappa=4.1$,
we see that the Kappa distribution carries $1-F_{\rm E}(x) \approx 19$\% of the entire energy in this tail of $v > 3\theta$. 
Since 1\% of the tail population is lost and redistributed,
one can estimate that $0.19$\% of the energy density is erroneously lost.
This is in good agreement with the energy loss of $\mathcal{O}(10^{-2.5})$
near $\kappa \approx 4.1$ in Figure~\ref{fig:DE}(b).

\begin{figure*}[htbp]
\centering
\includegraphics[width={0.9\textwidth}]{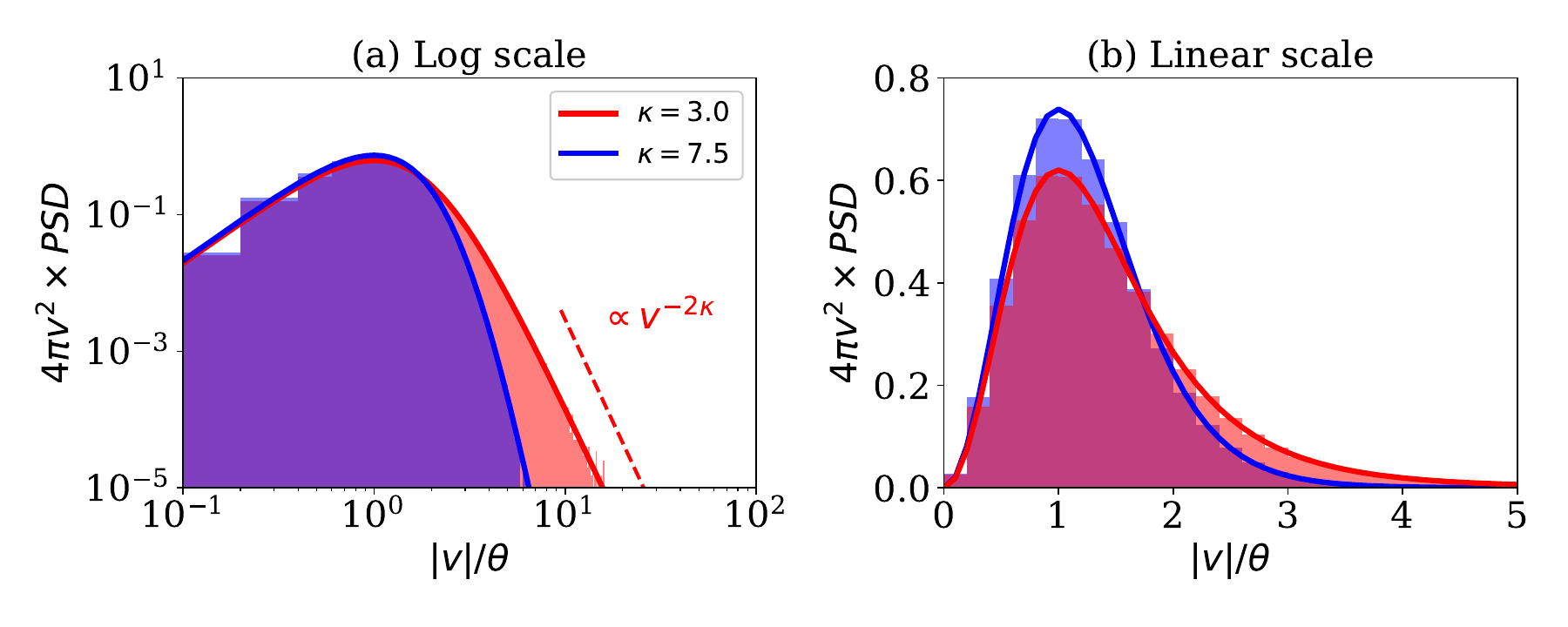}
\caption{
Monte Carlo generation of the approximated Kappa distribution (Eq.~\eqref{eq:approx}; histograms)
and the exact solutions (Eq.~\eqref{eq:kappa}) of the Kappa distribution (solid curves)
(a) on the logarithmic scale and
(b) on the linear scale.
\label{fig:histogram}}
\end{figure*}

\section{Random Number generator}
\label{sec:generator}

Next, we construct an RNG for
the approximate distribution, Eq.~\eqref{eq:approx}.
We first equate Eq.~\eqref{eq:G} and a uniform random variate $U_1 \sim {\rm Uniform}(0,1)$.
\begin{align}
\left\{ 1 - \exp_{q*}\left( -\frac{ax+bx^2}{1+cx} \right) \right\}^{3/2} = U_1
.\label{eq:GU}
\end{align}
With help from the $q$-logarithm,
the inverse function of the $q$-exponential function (Eq.~\eqref{eq:qexp}):
\begin{align}
\ln_q(x) \equiv \dfrac{ x^{1-q}-1 }{ 1-q }
,
\label{eq:qlog}
\end{align}
one can solve an inverse problem of Eq.~\eqref{eq:GU},
by way of a quadratic equation for $x$ \citep{umeda24}.
Then we obtain the particle velocity $v$ in the following way:
\begin{align}
v
&= \theta \sqrt{ \frac{-(a+cL) + \sqrt{(a+cL)^2 - 4bL} }{2b} }
= \theta \sqrt{ \frac{-2L}{(a+cL) + \sqrt{(a+cL)^2 - 4bL} } }
\label{eq:v}
\\
L &=
\ln_{q^*}\!\left( 1-U_1^{2/3}\right)
%= \dfrac{ \left( 1-U^{2/3}\right)^{1-{q^*}}-1 }{ 1-{q^*} }
= -\kappa^* \left\{ \left( 1-U_1^{2/3}\right)^{-1/\kappa^*} - 1 \right\}
\label{eq:L}
\end{align}
Note that $L$ is negative.
After obtaining $v$, we randomly scatter it in 3D directions
using the standard method with two uniform variates, $U_2$ and $U_3$.
All the procedure is presented
in the pseudocode in Table \ref{table:kappa}.
The first four lines are required only once, and 
the other lines need to be repeated for each particle.
Importantly, this algorithm just calls three uniform variates and
it has no loops or conditional statements.
All the variables are sequentially calculated.

\begin{table}
\centering
\caption{Numerical algorithm of the Kappa generator.
\label{table:kappa}}

\begin{tabular}{l}
\hline
$\kappa^* \leftarrow \kappa - 1/2$ \\
$a \leftarrow
\frac{1}{\kappa}
\left(\frac{2}{3 B(3/2,\kappa^*)}\right)^{2/3}$ \\
$c \leftarrow \dfrac{0.123 \kappa^{2} - 1.12\, \kappa + 2.56}{\kappa^{2} - 7.89\, \kappa + 15.6}$ \\
$b \leftarrow
\left\{ \kappa^* \frac{3}{2} B(3/2,\kappa^*)\right\}^{1/\kappa^*} \frac{\kappa^*}{\kappa} \,c$\\
\hline
generate $U_1, U_2, U_3 \sim {\rm Uniform}(0, 1)$ \\
$L \leftarrow -\kappa^* \left\{ \left( 1-U_1^{2/3}\right)^{-1/\kappa^*} - 1 \right\}
$ \\
$V \leftarrow \theta \sqrt{ \dfrac{-2L}{(a+cL) + \sqrt{(a+cL)^2 - 4bL} } }$ \\
$v_x \leftarrow V ( 2 U_2 - 1 )$ \\
$v_y \leftarrow 2 V \sqrt{ U_2 (1-U_2) } \cos(2\pi U_3)$ \\
$v_z \leftarrow 2 V \sqrt{ U_2 (1-U_2) } \sin(2\pi U_3)$ \\
%\textbf{return} $v_x, v_y, v_z$\\
\hline
\end{tabular}
\end{table}

\section{Numerical tests}
\label{sec:test}

To validate the proposed method in Table \ref{table:kappa},
we have carried out three numerical tests.
If necessary, we compare the proposed method
to two known methods:
(1) the standard Kappa generator, based on the $t$-distribution \citep{abdul15,zeni22},
and (2) the rejection method, based on the Pareto distribution \citep{zeni25}.
We call them the standard method and the Pareto method, respectively.
Inside the standard method,
we employ the \citet{mt00} method to generate a gamma variate.
The Pareto method has two options in its index $n$, but we choose a simple option $n=\kappa/2$, because it runs substantially faster than the other option.

The first test is intended for demonstration. 
Using the proposed method,
we have generated $10^6$ particles with $\kappa = 3.0$ and $\kappa=7.5$.
Their omni-directional velocity distributions are presented in the histograms in Figure \ref{fig:histogram}
on logarithmic and linear scales.
For example, in Figure \ref{fig:histogram}(a),
one can see that the $\kappa=3$ distribution has a power-law tail
with the index of $-2\kappa$.
For another example, in Figure \ref{fig:histogram}(b),
it is evident that the Kappa distributions have their maximum
at the most probable speed, $|v| = \theta$.
These are characteristic features of the Kappa distribution.
In both panels of Figure \ref{fig:histogram},
the histograms are in excellent agreement with the exact Kappa distributions,
even though the particle distributions actually follow the approximate distributions.

\begin{figure}[htbp]
\centering
\includegraphics[width={0.7\columnwidth}]{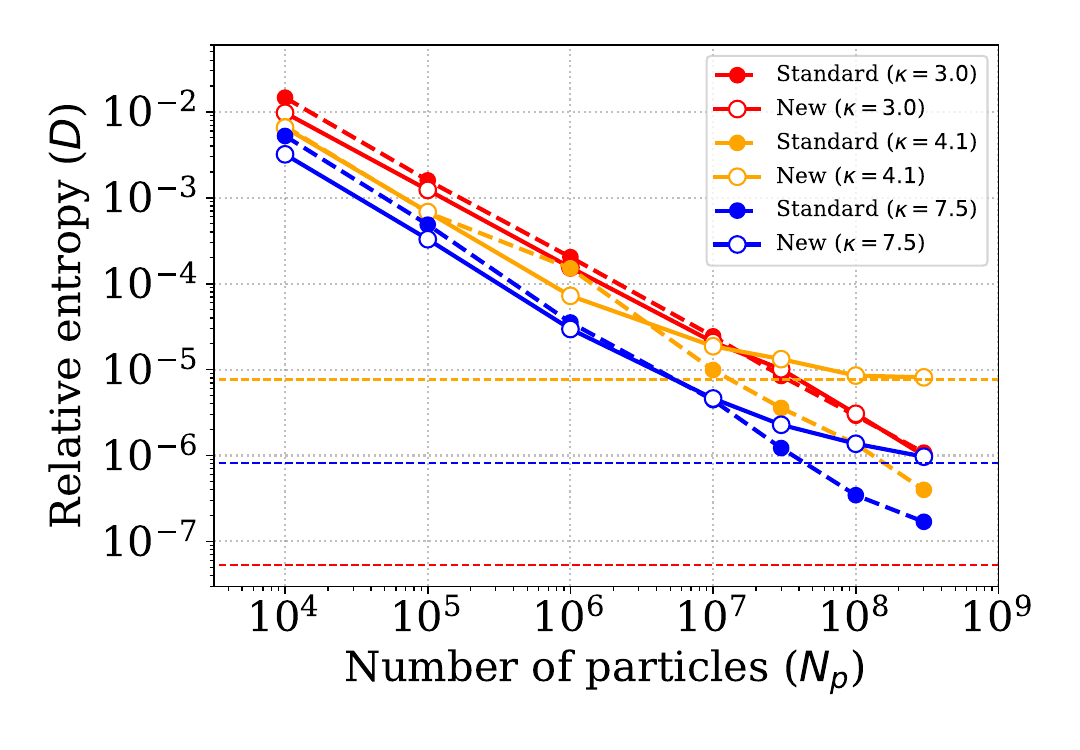}
\caption{
Relative entropy ($D$) between the numerically generated distribution and the exact solution for $\kappa=3.0, 4.1,$ and $7.5$, as a function of the number of particles ($N_p$).
\label{fig:KL}}
\end{figure}

Second, we evaluate the relative entropy
between the Kappa distribution and Monte Carlo results.
We consider the following form,
\begin{align}
D(P\|Q) = \sum_i P(i) \ln \frac{P(i)}{Q(i)}
\end{align}
where $P$ is the baseline distribution and
$Q$ is the target distribution.
For the baseline probabilities $P(i)$,
we numerically integrate the Kappa distribution (Eq.~\eqref{eq:kappax})
in each histogram bin with $\Delta v =1/5$.
For the target distributions $Q(i)$,
we use our Monte Carlo data in the same bin.
A small number $\epsilon=10^{-10}$ is added to avoid $\log 0$,
i.e.,
\begin{align}
D(P\|Q) \equiv \sum_i (P(i)+\epsilon) \ln \frac{P(i)+\epsilon}{Q(i)+\epsilon}
\label{eq:KL}
\end{align}
Note that we discussed $\mathcal{D}$ for continuous distributions in Eq.~\eqref{eq:relative_entropy}, while
$D$ is calculated from the discretized probabilities in Eq.~\eqref{eq:KL}.
They are essentially similar, $\mathcal{D} \simeq D$.
The finite width of the bins and the number $\varepsilon$ only make a small difference. 

Figure \ref{fig:KL} shows the relative entropy $D$
as a function of the total particle number
for $\kappa=3.0$ (red), $4.1$ (orange), and $7.5$ (blue).
Monte Carlo results are calculated by two methods.
The dashed lines indicate the results by the standard $t$-generator method \citep{abdul15,zeni22},
while the solid lines indicate those by the proposed method.
To clarify, the standard method
essentially draws random numbers from the exact Kappa distribution,
while the proposed method draws random numbers from the approximated distribution.

For $\kappa=3.0$,
the two methods give similar results.
The two relative entropies become smaller and smaller,
anti-proportional to the number of particles.
This suggests that the two random number distributions
converge toward the Kappa distribution. 
The proposed method gives excellent results,
because the approximated distribution is remarkably close to the Kappa distribution, as demonstrated in Figure \ref{fig:PSD}(b). 
For $\kappa=4.1$ and $7.5$,
the two methods start to deviate at $N_p \approx 10^7$ or $10^{7.5}$.
While the results by the standard method continue to scale beyond $N_p \gtrsim 10^7$,
the results by the proposed method are asymptotic to
$D \approx \mathcal{D}$ for the approximated distribution (Eq.~\eqref{eq:approx_x}). 
The asymptotic values are indicated by the dashed horizontal lines.
They correspond to the three circles in Figure \ref{fig:DE}(a).

Third, finally, we have evaluated 
the performance of the three Kappa generators:
(1) the standard method, (2) the Pareto method, and (3) the proposed method. 
We wrote our CPU codes in C, compiled them with
the {\tt clang} compiler (v18.1) with the {\tt -lgsl -lm} options,
and ran them on AMD Ryzen 5955 processor on Ubuntu Linux 24.04LTS.
Similarly, we wrote our GPU codes in CUDA, compiled them with the {\tt nvcc} compiler (v12.0), and ran them with an NVIDIA RTX A6000 GPU.
We carried out a series of runs to generate $10^8$ particles,
as a function of $\kappa$, ranging from 1.6 to 15. 
To completely rule out startup issues,
for each case, we ran the code four times, and then
averaged elapsed times of the last three runs.

\begin{figure*}[htbp]
\centering
\includegraphics[width={0.9\textwidth}]{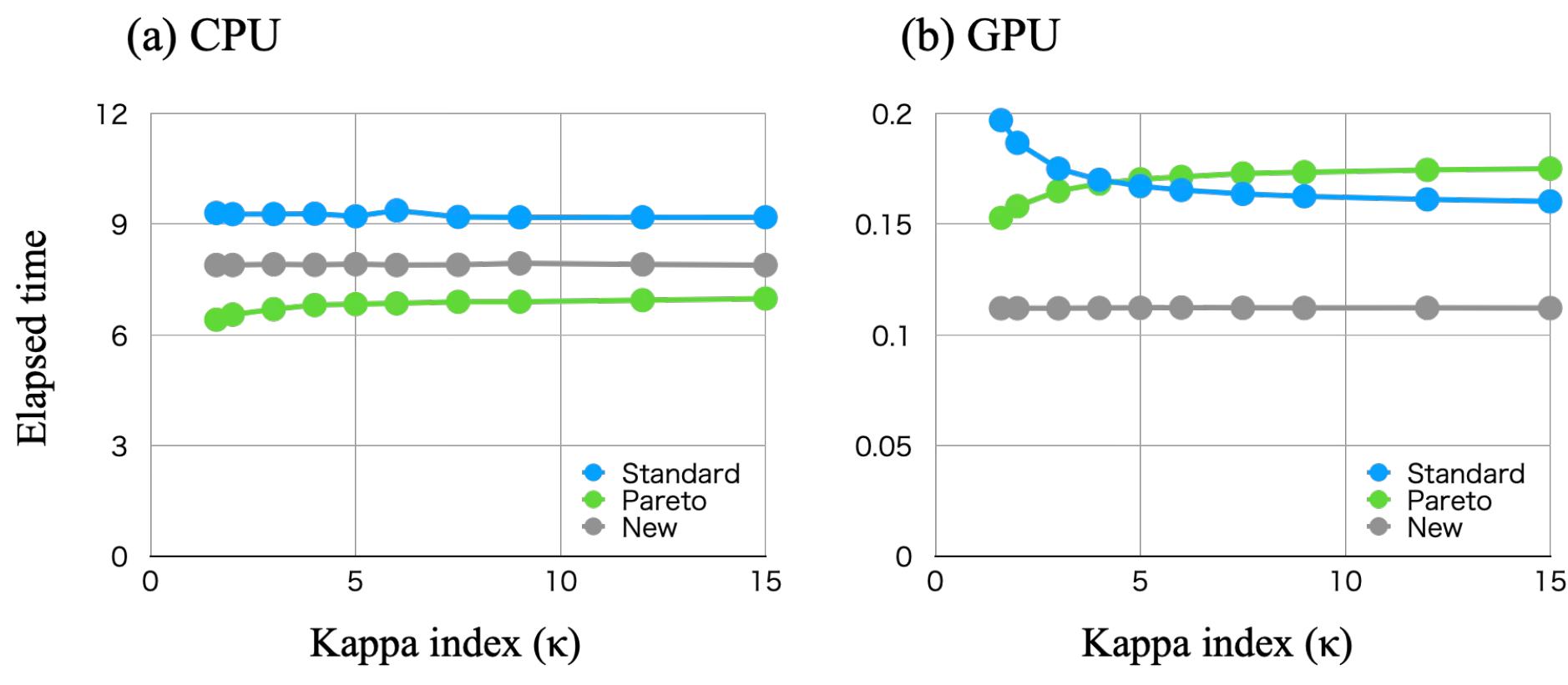}
\caption{
Elapsed times of the Kappa generators as a function of $\kappa$.
They are measured in seconds (a) on the CPU and (b) on the GPU.
\label{fig:benchmark}}
\end{figure*}

Figure \ref{fig:benchmark} shows the results.
On the CPU (Figure \ref{fig:benchmark}(a)),
we find that the proposed method (in gray) runs faster than
the standard method (in blue).
Furthermore, the Pareto method is the fastest,
possibly due to its simplicity.
The Pareto method shows a weak $\kappa$-dependent variation, 
while the other methods show only little variation.
On the GPU (Figure \ref{fig:benchmark}(b)),
the proposed method runs substantially faster than the other methods.
It essentially shows no variation, regardless of $\kappa$.
In contrast, the other methods show significant variation.
We attribute these trends to
the nature of the SIMT algorithms,
as discussed in the next section.

\section{Discussion and Summary}
\label{sec:discussion}

In the first and second tests,
we find that the proposed method provides a remarkably good approximation.
The results are visually indistinguishable from the exact solutions (Fig.~\ref{fig:histogram}),
as the difference around the most probable speed $v\approx\theta$ is kept small (Fig.~\ref{fig:PSD}(b)). 
In the low-$\kappa$ regime ($\kappa < 4$),
the proposed method gives a near-perfect approximation,
as highlighted in the low $\mathcal{D}$ values (Fig.~\ref{fig:DE}(a)).
The Monte Carlo results with $\kappa=3.0$ are indistinguishable
from the Kappa distribution in Fig.~\ref{fig:KL}.
Extrapolating the results toward the asymptotic value,
we will see a difference when $N_p \gtrsim 10^{9.5}$. 
In the high-$\kappa$ regime ($\kappa \gtrsim 5$),
the difference between the two distributions emerges
in the high-energy part (Fig.~\ref{fig:PSD}(b)).
Since phase space density in the high-energy part is
already very low for high-$\kappa$ cases,
the overall properties of the distribution are unlikely to be changed. 
Figures \ref{fig:DE}(a) and \ref{fig:DE}(b) tell us that
the discrepancy is most pronounced
in the transition case of $\kappa \approx 4.1$.
In Fig.~\ref{fig:KL}, the relative entropies deviate
when $N_p \gtrsim 10^{7.5}$ ($\kappa \gtrsim 5$) and
$N_p \gtrsim 10^{7}$ ($\kappa \approx 4.1$). 
Below these particle numbers, the difference between the approximate and exact Kappa distributions is completely hidden by the particle noise.
Since today's PIC simulation typically uses $10^2$--$10^3$ particles in a grid cell,
the proposed method would be accurate enough in many cases. 
The difference between the two distributions certainly emerges in $D$ beyond these numbers; however, even in the large-number limit, the PSD profile still looks similar to that of the Kappa distribution (Fig.~\ref{fig:PSD}(a)), at least on the logarithmic scale. 
Probably the biggest concern is
the total energy density (Fig.~\ref{fig:DE}(b)).
The normalized error is within $10^{-3}$ in both low- and high-$\kappa$ regimes,
but it increases to $\mathcal{O}(10^{-2.5})$ in the transition case. 
If the user wants to explore the energy balance at the accuracy of $\mathcal{O}(10^{-2.5})$, the proposed method is not ideal.

In the third test (Fig.~\ref{fig:benchmark}),
except for the Pareto method on the CPU,
the proposed method outperforms
the conventional methods on both CPU and GPU platforms,
particularly on the GPU.
We argue that the proposed method does not suffer from
the side effects of the rejection method on GPUs, unlike the conventional methods.
This is because the inverse transform method is SIMT-friendly,
while the rejection method is not.
In the following paragraphs, we discuss this side effect in more detail.

For simplicity, we estimate the computational costs
by using the total number of the uniform or normal variates per particle.
All the other factors are ignored. 
Table \ref{table:kappa} tells us that
the proposed method requires three uniform variates. 
The standard method calls three normal variates, and
two (one uniform and one normal) variates inside the gamma RNG,
which employs the \citet{mt00}'s rejection method (hereafter the MT method). 
The acceptance rate of the MT method is given by (\citet{yotsuji10}, eq. 3.308),
\begin{align}
p_1(k) =
\frac{ e^{k-1/3} \Gamma(k) }
{ \sqrt{2 \pi} (k-\frac{1}{3})^{k-1/2} }
\label{eq:mt_eff}
\end{align}
where $k=\kappa-1/2$ is the shape parameter of the gamma distribution.
The rate increases from $0.95$ to $1.0$,
as $\kappa$ ranges from $3/2$ to infinity.
To summarize,
the standard Kappa generator requires $3+2/p_1$ variates per particle on CPUs. 
The Pareto method is a simple rejection method.
It requires two uniform variates outside the rejection loop,
another two inside the loop, and
its acceptance rate is \citep{zeni25}
\begin{align}
p_2(\kappa)
=
\frac{\kappa^{1+\kappa/2}}{2 \sqrt{ \left( {\kappa-1} \right)^{\kappa - 1} }}
B\left( \frac{3}{2},\kappa - \frac{1}{2} \right)
.\label{eq:pareto_eff}
\end{align}
The rate ranges from $0.81$ to $0.73$.
Considering these, the Pareto method requires $2+2/p_2$ variates on CPUs.

\begin{figure}[htbp]
\centering
\includegraphics[width={0.6\columnwidth}]{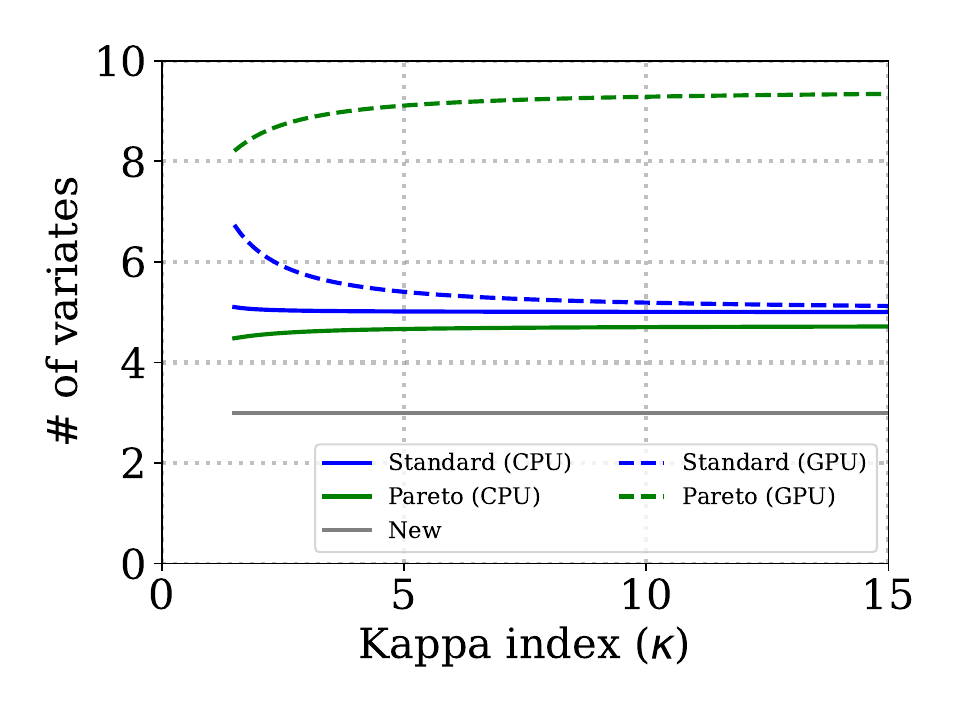}
\caption{
Estimated number of random variates per particle
in the Kappa generators.
\label{fig:variates}}
\end{figure}

The situation is different on GPUs.
Inside GPUs, typically
32 threads are controlled by the same instruction.
In such a case, one can proceed to the next computational step,
only after all 32 threads complete the loop. 
According to \citet{ridley21},
the typical count $\mu$ of the rejection loop is approximated by the mode of the exponentiated geometric distribution \citep{cg15}.
Depending on the rejection rate at each step, $\rho = 1-p$, and
the number of SIMT threads $t$, $\mu$ is approximated by
\begin{align}
\mu_1(\rho,t)
&\approx
3 - (1-\rho)^t - (1-\rho^2)^t
~~&(\rho \lesssim 2^{1/t}-1)
\label{eq:ridley1}
\\
\mu_2(\rho,t)
&\approx
\dfrac{\log \left( \frac{2(\rho-1)t}{\log(\rho)} \right) }{\log(1/\rho)} + 1
&(\rho \gg 2^{1/t}-1)
\label{eq:ridley2}
\end{align}
For $t=32$, the threshold value $2^{1/t}-1$ is $\approx 0.022$.
In the case of the standard method,
Eq.~\eqref{eq:mt_eff} gives the rejection rate of
$\rho = 1-p_1(\kappa-1/2) \lesssim 0.05$.
This sometimes exceeds the threshold, however,
Eq.~\eqref{eq:ridley1} actually gives a better approximation than Eq.~\eqref{eq:ridley2} for $\rho \lesssim 0.1$,
as shown in Figure 6 in \citet{ridley21}. %and
%in the supporting material \citep[][the last cell in the notebook]{zeni26}. 
Thus, we estimate the average number of variates to be $3+2\mu_1(1-p_1,32)$ on GPUs.
On the other hand, the Pareto method (Eq.~\eqref{eq:pareto_eff}) gives
the rejection rate of $0.19 < (1-p_2) < 0.27$,
which is greater than the threshold by an order-of-magnitude.
Eq.~\eqref{eq:ridley2} gives a better approximation in this case.
Although it is difficult to see, we can confirm this in Figure 5 in \citet{ridley21}. % and in the supporting material.
Then the number of variates is estimated to be $2+2\mu_2(1-p_2,32)$ on GPUs.
All these numbers are presented in Figure \ref{fig:variates}
as a function of $\kappa$.

Although crudely, Figure \ref{fig:variates} captures
the key trends in Figure \ref{fig:benchmark}.
The Pareto method becomes substantially slower on GPUs,
likely because
the rejection loop takes much more iterations on GPUs than on CPUs.
Their weak variations on $\kappa$ are also similar.
These results further support our argument that
these are influenced by the side effects on GPUs.
The standard method also runs slower on GPUs,
as $\kappa$ approaches $3/2$.
We recognize a sharper increase of the elapsed time near $\kappa=1.5$,
both in Figure \ref{fig:benchmark}(b) and in Figure \ref{fig:variates}. 
%Probably Figure \ref{fig:variates} underestimates the side effect,
%because the MT method contains some calculations inside the rejection loop. 
We do not see a slowdown in the proposed method.
Its behavior is independent of $\kappa$.
Since the proposed method does not contain the rejection loop,
it does not suffer from the slowdown in SIMT processors. 
In the future, we may need to use accelerators that employ more SIMT threads than 32.
In such a case, the conventional methods perform even worse,
while the proposed method should maintain its performance.

Practically, the proposed method is easy to implement.
The complete procedure is shown in Table \ref{table:kappa}.
In case the beta function is unavailable,
one can still calculate $B(a,b)$ using the gamma function:
$B(a,b)=\Gamma(a)\Gamma(b)/\Gamma(a+b)$.
One potential problem is that
a uniform variate is often generated in $\in (0,1]$ by GPU libraries
such as {\tt cuRAND} and {\tt hipRAND}.
In such a case, $U_1^{2/3}$ should be replaced by $(1-U_1)^{2/3}$ in Eq.~\eqref{eq:L},
to avoid a negative power of zero. 
Aside from these, we do not find serious technical concerns.

In summary, we have proposed a simple inverse transform method
for generating the Kappa distribution in particle simulations.
The method is fully described in Table \ref{table:kappa}.
It generates an approximate distribution,
which is particularly close to the Kappa distribution in the low-$\kappa$ regime. 
The relations between the particle number, accuracy, and $\kappa$ parameters are discussed.
Computationally, the new method outperforms the conventional methods in many cases on the CPU and on the GPU.
By definition, the proposed method is GPU-friendly,
as it is free from the side effects in the SIMT model.
We hope that the proposed method will be useful on future GPUs with more SIMT threads.

\section*{Acknowledgements}
TU was supported by Grant-in-Aid for Scientific Research (B) 24K00603 from the Japan Society for the Promotion of Science (JSPS).
The Jupyter notebook and C/CUDA codes are archived in Zenodo \citep{zeni26}.

\appendix
\section{Taylor expansions}
Here we formally derive Eqs.~\eqref{eq:F_taylor_0},
\eqref{eq:G_taylor_0}, \eqref{eq:F_taylor_8}, and \eqref{eq:G_taylor_8}.
We first evaluate the exact CDF of Eq.~\eqref{eq:CDF}.
For convenience, we use a new variable $y \equiv x/(x+\kappa)$.
Near $x=0$, we find
\begin{align}
y &= \frac{x}{\kappa} - \left( \frac{x}{\kappa} \right)^2 + \mathcal{O}(x^3)
\label{eq:appendix_y}
\\
y^{3/2}
&= \left(\frac{x}{\kappa}\right)^{3/2} \left(1-\frac{x}{\kappa}\right)^{3/2}
= \left(\frac{x}{\kappa}\right)^{3/2} \left( 1- \frac{3}{2\kappa} x + \mathcal{O}(x^2)\right)
\label{eq:appendix_f03}
\end{align}
Considering the definition of the incomplete beta function,
we expand Eq.~\eqref{eq:CDF} as follows:
%\begin{align}
%F(x)
%&= 
%%I_{\frac{x}{x+\kappa}}\left( \frac{3}{2}, \kappa-\frac{1}{2} \right)
%%= 
%\frac{B_y\left( \frac{3}{2}, \kappa-\frac{1}{2} \right)
%}{B\left( \frac{3}{2}, \kappa-\frac{1}{2} \right)
%}
%\end{align}
\begin{align}
F(x)
&= 
I_{y}\left( \frac{3}{2}, \kappa-\frac{1}{2} \right)
= 
\frac{1}{B\left( \frac{3}{2}, \kappa-\frac{1}{2} \right)}
\int_0^y t^{1/2} (1-t)^{\kappa-3/2} dt
\nonumber \\
&= 
\frac{1}{B\left( \frac{3}{2}, \kappa-\frac{1}{2} \right)}
\int_0^y t^{1/2}
\bigg( 1 - (\kappa-3/2)t + \mathcal{O}(t^2) \bigg)
dt
\nonumber \\
&= 
\frac{y^{3/2}}{\frac{3}{2} B\left( \frac{3}{2}, \kappa-\frac{1}{2} \right)}
\bigg( 1 + \frac{3({3/2-\kappa})}{5}  y + \mathcal{O}(y^2) \bigg)
\label{eq:appendix_f0}
\end{align}
We can also verify this expansion,
with help from
the hypergeometric function ${}_2F_1(a,b;c;x)$ and
formulas in \citet[][8.1.7, 15.2.1]{dlmf},
\begin{align}
F(x)
&= 
\frac{y^{3/2}
}{ \frac{3}{2} B\left( \frac{3}{2}, \kappa-\frac{1}{2} \right)
}
{}_2F_1\left( \frac{3}{2}, \frac{3}{2}-\kappa; \frac{5}{2} ;  y \right)
= 
\frac{y^{3/2}
}{ \frac{3}{2} B\left( \frac{3}{2}, \kappa-\frac{1}{2} \right)
}
\left( 1 + \frac{3(3/2-\kappa)}{5} y + \mathcal{O}(y^2) \right)
\end{align}
Substituting Eqs.~\eqref{eq:appendix_y} and \eqref{eq:appendix_f03}
into Eq.~\eqref{eq:appendix_f0}, and since $y^2$ is an order of $\mathcal{O}(x^2)$ ,
we find 
\begin{align}
F(x)
&= 
\frac{2}{ 3 B\left( \frac{3}{2}, \kappa-\frac{1}{2} \right)}
\left(\frac{x}{\kappa}\right)^{3/2}
\left( 1 + \frac{3(\frac{3}{2}-\kappa)}{5} \frac{x}{\kappa} + \mathcal{O}(x^2) \right)
\left( 1- \frac{3}{2\kappa} x + \mathcal{O}(x^2) \right)
\nonumber \\
&=
\frac{2}{3B\left( \frac{3}{2}, \kappa-\frac{1}{2} \right)}
\left(\frac{x}{\kappa}\right)^{3/2}
\left(
1
-
\frac{3(\kappa+1)}{5\kappa}x
+
\mathcal{O}(x^2)
\right)
\end{align}
in agreement with Eq.~\eqref{eq:F_taylor_0}.

Next we examine the $x \rightarrow \infty$ limit.
In this case, we expand a new variable $z$ with respect to $(1/x)$.
\begin{align}
z
&\equiv 1 - y = {\kappa}{x}^{-1} - {\kappa^2}{x}^{-2} + \mathcal{O}(x^{-3})
\label{eq:appendix_z}
\\
z^{\kappa-\frac{1}{2}}
&= \left(\frac{\kappa}{x}\right)^{\kappa-\frac{1}{2}} \left(1 - \frac{\kappa}{x} \right)^{\kappa-\frac{1}{2}}
= \left(\frac{\kappa}{x}\right)^{\kappa-\frac{1}{2}} \left( 1- \kappa \left(\kappa-\frac{1}{2} \right){x^{-1}} + \mathcal{O}(x^{-2})\right)
\label{eq:appendix_f83}
\end{align}
Using the relation $I_x(a,b) = 1 - I_{1-x}(b,a)$, we find
\begin{align}
F(x)
&= 
1 - I_{z}\left( \kappa-\frac{1}{2}, \frac{3}{2} \right)
=
1 - \frac{1}{B\left( \frac{3}{2}, \kappa-\frac{1}{2} \right)
}
\int_0^z t^{\kappa-{3}/{2}} (1-t)^{1/2} dt
\nonumber \\
&= 
1 - \frac{1}{B\left( \frac{3}{2}, \kappa-\frac{1}{2} \right)
}
\int_0^z t^{\kappa-{3}/{2}} \left( 1 - \frac{1}{2}t + \mathcal{O}(t^2) \right) dt
\nonumber \\
&= 
1 -
\frac{z^{\kappa-\frac{1}{2}}
}{ \left( \kappa-\frac{1}{2} \right) B\left( \frac{3}{2}, \kappa-\frac{1}{2} \right)
}
\left( 1 - \frac{ \kappa-\frac{1}{2} }{2\kappa+1} z + \mathcal{O}(z^2) \right)
\label{eq:appendix_f8}
\end{align}
We can similarly check the results using the formulas
\citep[][8.1.7, 15.2.1]{dlmf},
\begin{align}
F(x)
&= 
%1 - \frac{B_{z}\left( \kappa-\frac{1}{2}, \frac{3}{2} \right)
%}{B\left( \frac{3}{2}, \kappa-\frac{1}{2} \right)
%}
%= 
1 -
\frac{z^{\kappa-\frac{1}{2}}
}{ \left( \kappa-\frac{1}{2} \right) B\left( \frac{3}{2}, \kappa-\frac{1}{2} \right)
}
{}_2F_1\left( \kappa-\frac{1}{2}, -\frac{1}{2}; \kappa+\frac{1}{2}; z \right)
\nonumber \\
&= 
1 -
\frac{z^{\kappa-\frac{1}{2}}
}{ \left( \kappa-\frac{1}{2} \right) B\left( \frac{3}{2}, \kappa-\frac{1}{2} \right)
}
\left( 1 - \frac{ \kappa-\frac{1}{2} }{2\kappa+1} z + \mathcal{O}(z^2) \right)
\end{align}
Substituting Eqs.~\eqref{eq:appendix_z} and \eqref{eq:appendix_f83}
into Eq.~\eqref{eq:appendix_f8}, and since $z^2$ is an order of $\mathcal{O}(x^{-2})$ ,
we find
\begin{align}
F(x)
&= 
1 -
\frac{\kappa^{\kappa-1/2}}{ \left( \kappa-\frac{1}{2} \right) B\left( \frac{3}{2}, \kappa-\frac{1}{2} \right)
}
\left( 1 - \frac{ \kappa-\frac{1}{2} }{2\kappa+1} \kappa x^{-1} + \mathcal{O}(x^{-2}) \right)
\left( 1 - \left(\kappa-\frac{1}{2} \right){\kappa x^{-1}} + \mathcal{O}(x^{-2})\right)
x^{-(\kappa-1/2)}
\nonumber \\
&=
1 - 
\frac{\kappa^{\kappa-1/2}}{ \left( \kappa-\frac{1}{2} \right) B\left( \frac{3}{2},\kappa - \frac{1}{2} \right)}
x^{-(\kappa-1/2)}
\left(
1
- 
\frac{ \left( \kappa-\frac{1}{2} \right) (\kappa+1) \kappa}{ \kappa+\frac{1}{2} }
{ x^{-1}}
+ 
\mathcal{O}(x^{-2})
\right)
\end{align}
This is in agreement with Eq.~\eqref{eq:F_taylor_8}.

Next we evaluate the approximate CDF of Eq.~\eqref{eq:G}.
We rewrite it as follows,
\begin{align}
G(x) =
\left\{ 1 - \bigg( 1 + \frac{1}{\kappa^*} \frac{ax+bx^2}{1+cx} \bigg)^{-\kappa^*} \right\}^{3/2}
.
\label{eq:appendix_g}
\end{align}
%We expand the fraction part,
%\begin{align}
%\frac{ax+bx^2}{1+cx} = ax + (b-ac)x^2 + \mathcal{O}(x^3)
%\end{align}
We expand the $q$-exponential part
\begin{align}
\bigg( 1 + \frac{1}{\kappa^*} \frac{ax+bx^2}{1+cx} \bigg)^{-\kappa^*}
&=
\bigg( 1 + \frac{1}{\kappa^*} \left\{ ax + (b-ac)x^2 + \mathcal{O}(x^3) \right\} \bigg)^{-\kappa^*}
\nonumber \\
&=
1 - 
\bigg( ax + (b-ac)x^2 + \mathcal{O}(x^3) \bigg)
+
\frac{(\kappa^*+1)}{2\kappa^*}
\bigg( ax + (b-ac)x^2 + \mathcal{O}(x^3) \bigg)^2
\nonumber \\
&=
1 - ax
+
\bigg(
\frac{(\kappa^*+1)a^2}{2\kappa^*}
- (b-ac)
\bigg)
x^2
+ \mathcal{O}(x^3) 
\label{eq:appendix_g03}
\end{align}
Substituting Eq.~\eqref{eq:appendix_g03} into Eq.~\eqref{eq:appendix_g},
we obtain Eq.~\eqref{eq:G_taylor_0}
\begin{align}
G(x) &=
\left\{ ax
+
\bigg(
(b-ac)
- \frac{(\kappa^*+1)a^2}{2\kappa^*}
\bigg)
x^2
+ \mathcal{O}(x^3) 
 \right\}^{3/2}
\nonumber \\
&=
a^{3/2}
x^{3/2}
\left\{
1 +
\bigg(
\frac{b-ac}{a}
- \frac{(\kappa^*+1)a}{2\kappa^*}
\bigg)
x
+ \mathcal{O}(x^2)
 \right\}^{3/2}
\nonumber \\
&=
a^{3/2}
x^{3/2}
\left\{
1
+
\frac{3}{2}
\left(
\frac{b-ac}{a}
-
\frac{(\kappa^*+1)a}{2\kappa^*}
\right)
x
+ \mathcal{O}(x^2)
\right\}
\end{align}

For $x\rightarrow\infty$, the $q$-exponential part yields
%we expand the fraction part with respect to $(1/x)$:
%\begin{align}
%\frac{ax+bx^2}{1+cx} = x \left( \frac{b}{c} + \frac{ac-b}{c^2} x^{-1} + \mathcal{O}(x^{-2}) \right)
%\end{align}
%Using this, the $q$-exponential function yields
\begin{align}
\bigg( 1 + \frac{1}{\kappa^*} \frac{ax+bx^2}{1+cx} \bigg)^{-\kappa^*}
&=
\bigg( 1 + \frac{x}{\kappa^*} \left\{ \frac{b}{c} + \frac{ac-b}{c^2} x^{-1} + \mathcal{O}(x^{-2}) \right\} \bigg)^{-\kappa^*}
\nonumber \\
&=
\bigg( \frac{c\kappa^*}{bx} \bigg)^{\kappa^*}
\bigg( 1 + \frac{c}{b}\left\{ \kappa^* + \frac{ac-b}{c^2} \right\} x^{-1} + \mathcal{O}(x^{-2}) \bigg)^{-\kappa^*}
\nonumber \\
&=
\bigg( \frac{c\kappa^*}{bx} \bigg)^{\kappa^*}
\bigg( 1 - \frac{c\kappa^*}{b}\left\{ \kappa^* + \frac{ac-b}{c^2} \right\} x^{-1} + \mathcal{O}(x^{-2}) \bigg)
\label{eq:appendix_g83}
\end{align}
Substituting Eq.~\eqref{eq:appendix_g83} into Eq.~\eqref{eq:appendix_g},
we arrive at Eq.~\eqref{eq:G_taylor_8}
\begin{align}
G(x) &=
\left\{ 1
-
\bigg( \frac{c\kappa^*}{b} \bigg)^{\kappa^*} x^{-\kappa^*}
\bigg( 1 + \frac{c\kappa^*}{b}\left\{ \kappa^* + \frac{ac-b}{c^2} \right\} x^{-1} + \mathcal{O}(x^{-2}) \bigg)
\right\}^{3/2}
\nonumber \\
&=
1
-
\frac{3}{2}
\left(\frac{c \kappa^*}{b}\right)^{\kappa^*}
x^{-\kappa^*}
+
\mathcal{O}(x^{-(\kappa^*+1)})
\end{align}

\end{document}